\documentclass{PoS}

\def\LAPPDTM{LAPPD\textsuperscript{TM}~}

\title{ Drifting Photons on Optical Paths, Mirrors, Sub-mm
  Resolution in Four Dimensions,  and Transverse/Longitudinal Phase Space:
  Exploiting Psec Time Resolution}

\ShortTitle{Drifting Photons on Optical Paths with Mirrors and Timing}

\author{\speaker{Henry J. Frisch}\\
        Enrico Fermi Institute and Department of Physics\\
        University of Chicago
        E-mail: \email{frisch@hep.uchicago.edu}}

\abstract{I discuss the status of MCP-based photo-detector
amplification sections and Cherenkov light sources for precise
timing measurements of charged particles and gamma rays. Sub-psec
resolution is predicted for the large pulses such as those produced
by a charged particle or electromagnetic shower traversing a
photo-detector entrance window. Measuring events with sub-mm
resolution in each of the 4 dimensions expands the optical phase
space from 4 dimensions, allowing emittance transformations that
can minimize expensive instrumented photo-sensitive area.}

\FullConference{5th International Conference on Micro-Pattern Gas Detectors (MPGD2017)\\
        22-26 May, 2017\\
        Philadelphia, USA}

\begin{document}
This is a talk given to a sophisticated audience of particle
detector developers who understand the many challenges of
inventing and bringing into routine use an innovative detector
technology on a large scale. One can ask why a talk on very fast
optical vacuum-tube detectors in a conference on gas-based
detectors? I hope what follows elicits ideas for novel hybrid
 detector systems as well as for improving and
exploiting timing in event reconstruction in gas-based devices.
%\section{}
\section{Three Timing Cases to Distinguish}
The factors limiting the ultimate timing resolution are different in
each of the following cases:
\begin{enumerate}
\setlength{\itemsep}{-0.03in}
\item Single optical photons (Scintillation or Cherenkov)
\item Charged particles above Cherenkov threshold (in $H_2O$ or a
  glass window, for example)
\item Electromagnetic showers from high energy photons.
\end{enumerate}

 I will talk first about the 2nd and 3rd
cases, relativistic charged particles and high-energy photons, for
which psec or sub-psec time resolutions are plausible given that
certain detection criteria are met, before moving on to the 1st
case, for which the ultimate resolutions are determined by other
factors and are typically an order-of-magnitude larger.

I treat time and space distance in `natural' units, i.e. $c=1$,
and 1 psec $\simeq$ 300 microns; 1 nsec $\simeq$ 1 foot. Similarly
momentum is defined as $pc$ (and so has the same units as
energy~\footnote{As $(ct,\vec{x})$ and $(E,c\vec{p})$ are
    Lorentz 4-vectors it is natural to use the same units for all four
  components when discussing light (and unnatural to do otherwise).}.

\section{Criteria for Sub-Psec Timing}
\begin{figure}[h]
\includegraphics[width=.6\textwidth]{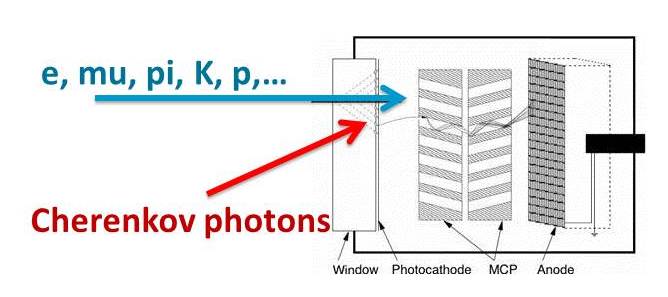}
\caption{The first psec criterion: a fast bright source. A prime
example is the Cherenkov light from a charged particle traversing
a radiator or the entrance window of a photodetector as shown.}
\label{fig:cherenk_in_window}
\end{figure}

The first necessary criterion for sub-psec timing is a 
  source of many photons in a psec time-space interval. A prime
  example is the Cherenkov light from a charged particle traversing a
  radiator or the entrance window of a
  photodetector~\cite{Credo,Ohshima,Vavra_Fermilab_paper,Anatoly_paper}.
  Figure~\ref{fig:cherenk_in_window} shows the geometry.
  The first light to hit the photocathode on the
   vacuum side of the window will be at the spot the particle
   exits the window, followed by photons in an expanding ring
   as the Cherenkov light from the path arrives at the window exit plane
   (assuming normal incidence- the ring shape and timing depend on  the
    incident angle of the charged particle). A fast (coherent)  pulse of photons
    impinging on the cathode will produce many photo-electrons in a well-defined pattern.

\begin{figure}
\includegraphics[width=.6\textwidth]{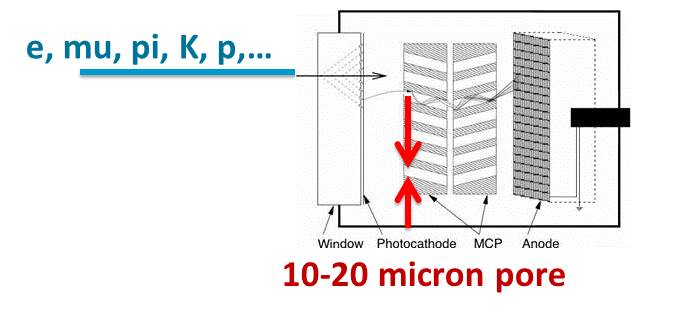}
\caption{The second psec criterion: a homogeneous array of pixels
small compared to a psec.}
  \label{fig:small_pores}
\end{figure}

The second criterion is a homogeneous array of small pixels so
that light transit times are short within one pixel. The small
pixels produce a fast rise-time such as in a Micro-Channel Plate
(MCP)~\cite{Photek}. An Incom~\cite{Incom} 8''-square
micro-channel capillary plate has 8$\times 10^7$ 20-micron pores
(20 microns is 67 femtoseconds at $c$).
Figure~\ref{fig:small_pores} indicates the geometry.

\begin{figure}[h]
\includegraphics[width=.6\textwidth]{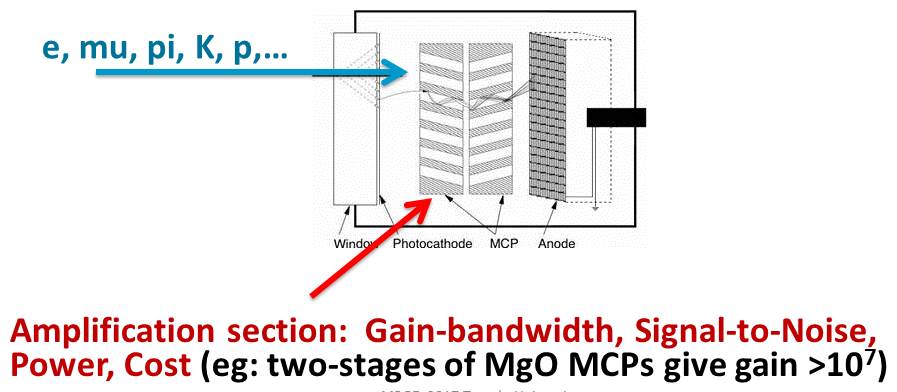}
\caption{The third psec criterion: high-gain per pixel., so that
the  {\it first} photo-electron determines the time of the pulse.}
\label{fig:high_gain}
\end{figure}

The third psec criterion: high-gain per pixel, is  more subtle.
The bright source
  will produce multiple photons, typically with a transit-time-spread
  (TTS) in the range of 50-100 picoseconds. However the {\it first} photon
  will have a much smaller jitter. High gain is necessary for the
  first photon to determine the timing of the subsequent pulse. The statistics
  determining the timing consequently is not Gaussian,
but Poisson; the determining factor is the probability for an
elapsed time from the particle traversal in which there is {\it
no} photoelectron, which per unit time is $e^{-\bar{n}}$, where
${\bar{n}}$ is the average number of photoelectrons per psec.
Figure~\ref{fig:high_gain} depicts the idea. A full simulation
will be required to make this idea quantitative (See
Figure~\ref{fig:pore_simulation}).


\newpage
\section{Factors that Determine Timing Resolution and the 1 Psec `Barrier'}

The LAPPD Collaboration was formed in 2009 to build large-area
fast detectors for charged particle identification at the Fermilab
Tevatron Collider and Large Hadron Collider at
CERN~\cite{History_paper}.  At the first of the workshops on
determining the limits in fast timing~\cite{PSEC_library} there
was spirited debate on the most promising techniques and the
factors that limit the resolution. Subsequent development of
electronics and photodetectors, and detailed
measurements~\cite{Doc_lib}  have pushed the uncertain region down
to $~5$ psec (See, for example, the contributions to {\it The
Factors That Limit Time Resolution}~\cite{limitations_workshop}).

\begin{figure}
\includegraphics[width=.7\textwidth]{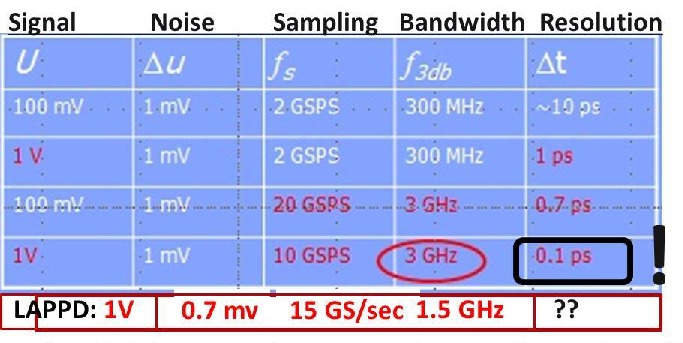}
\caption{A table from Stefan Ritt's talk at the Second Chicago
  Photocathode Workshop characterizing the predicted timing resolution
  using waveform sampling, For a fixed number of samples on the
  leading edge of the pulse the dependence on the Signal-to-Noise
  ratio ($U/\Delta U$) and bandwidth is linear. The current measured
  $\LAPPDTM$ values are shown in the last row.}
\label{fig:annotated_Ritt_table}
\end{figure}

Figure~\ref{fig:annotated_Ritt_table} shows a table from Stefan Ritt's
talk at the Second Chicago Photocathode Workshop characterizing the
predicted timing resolution using waveform
sampling~\cite{Ritt_workshop} . For a fixed number of samples on the
leading edge of the pulse the dependences on Signal/Noise ($U/\Delta
U$) and bandwidth are linear. Current $\LAPPDTM$ values are shown in
the last row~\cite{LAPPDTM}. This is of course an extrapolation by
more than an order of magnitude. However as I will show below, LAPPD
data follow the rule-of-thumb down to $~6$ psec and extrapolate to 1-2
psec.

\begin{figure}
\includegraphics[width=.50\textwidth]{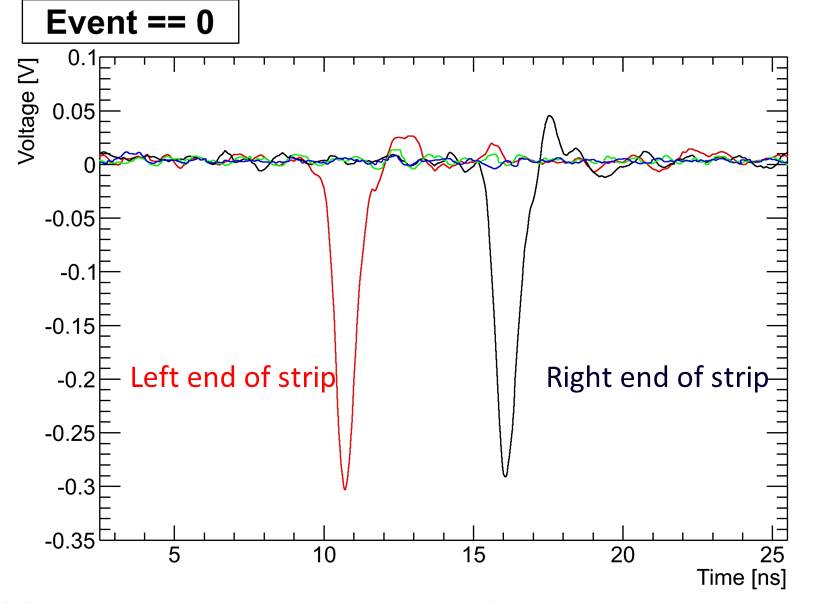}
\caption{A pulse digitized at the two ends of one of the 50-ohm anode strips
  measured in a pumped (i.e. not hermetically sealed) LAPPD detector at
  the ANL-APS laser lab~\cite{RSI_paper}.}
\label{fig:pulses_strip_ends}
\end{figure}

Figure~\ref{fig:pulses_strip_ends} shows the anode pulse resulting
from a fast laser pulse
 incident on the `Demountable' \LAPPDTM~\cite{Timing_paper}. The photodetector anode
 consists of 50-ohm RF strips~\cite{Tang_Naxos}
 which penetrate the hermetic package of the photodetector.
 The strips  are read out at 10 GS/s with waveform sampling ASICs at each
end of each strip. The time of the laser pulse is given by the
average time of the two recorded pulses; the position by the
difference in times.

\newpage
\section{Waveform Sampling: Oscilloscopes on Chips}
\begin{figure}
\includegraphics[width=.6\textwidth]{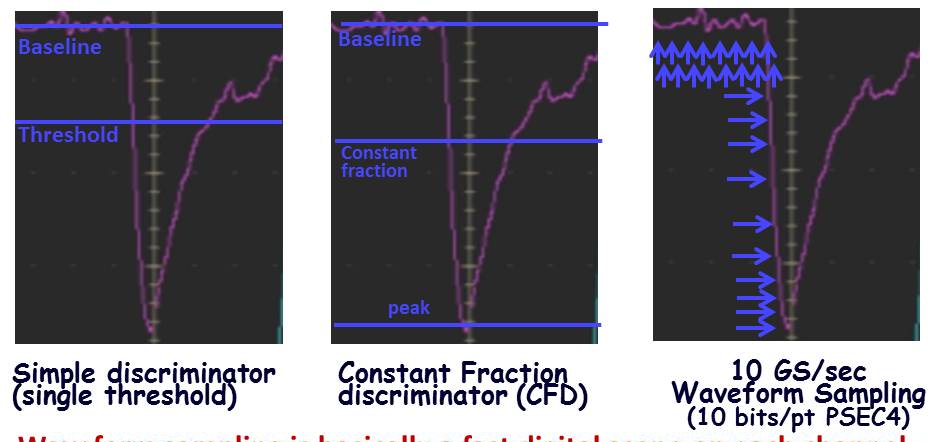}
\caption{A graphic comparing waveform sampling to two work-horse
methods of determining the time of a fast pulse: single threshold
and constant-fraction discriminator. }
\label{fig:timing_sampling_cfd_disc_comparison}
\end{figure}

The development of psec timing in real experiments in the field has to take
into account systematic effects such as pile-up,  baseline shifts due to lower-frequency noise,
 photodetector light emission and HV breakdown, and pulse slewing.
The development of waveform sampling
ASICS~\cite{waveform_sampling} leads to having the equivalent of a
fast oscilloscope on each channel, and, as shown in
Figure~\ref{fig:timing_sampling_cfd_disc_comparison}, is
inherently more precise than a simple threshold or a Constant
Fraction Discriminator~\cite{JF_NIM}.

\begin{figure}
\includegraphics[width=.6\textwidth]{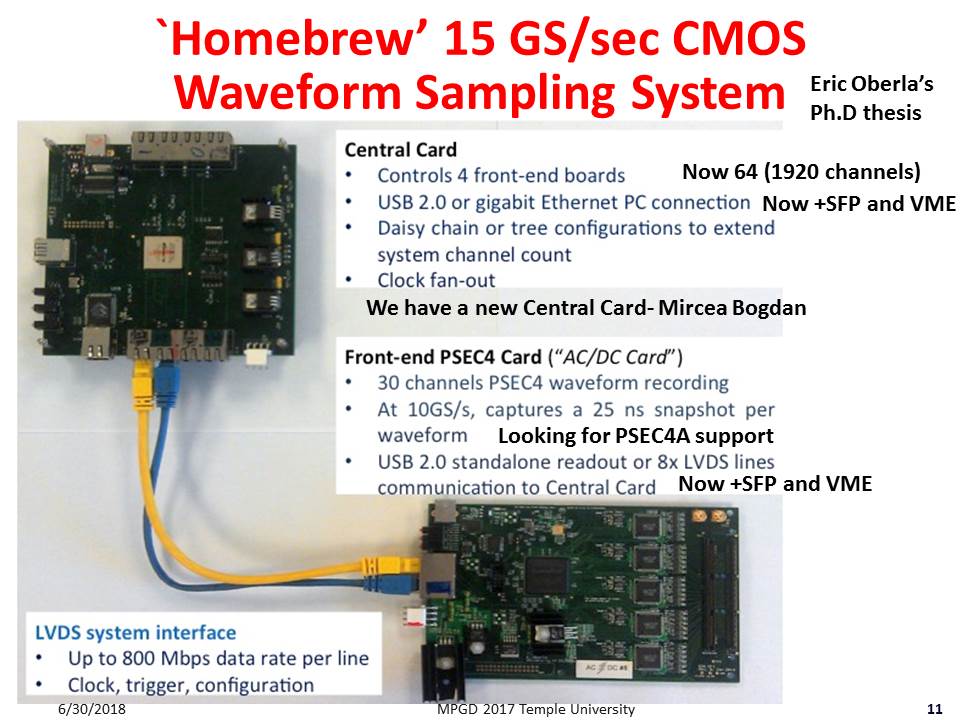}
\caption{The ACC control card (upper left) and the current ACDC
front-end card (lower right).}
\label{fig:ACDC_Controlcard_system}
\end{figure}

%YOUAREHERE
% Figure 7
Figure~\ref{fig:ACDC_Controlcard_system} shows the structure of
the UC PSEC4-based 10-15 Gs/sec waveform sampling
  system~\cite{PSEC4_paper,Mircea_Strasbourg}.
  The ACC control card (top left) can support up to 1920 channels in a
  VME crate. The current ACDC front-end card supports 30 channels of PSEC4
  waveform recording at 10-15 GS/s at better than 10 bits. The
  Achilles heel of this version of the PSEC4 ASIC is the short buffer of 256
  samples.

\section{Test-bench Measurements with the LAPPD Photodetector Prototypes}

The LAPPD effort~\cite{History_paper} can be characterized as five
interlinked but remarkably independent development projects: 1) a
high-gain low-ion-feedback ALD-coated amplification subsystem
consisting of a chevron of 20-cm MCPs; 2) a high-bandwidth (GHz)
segmented anode capable of sub-mm resolution; 3) large-area
bialkali photocathodes; 4) an inexpensive robust low-mass
low-profile hermetic package; and 5) multi-channel low-power
electronics readout capable of psec resolution. The first, second,
and fifth of these lend themselves to testing `on the bench' in a
well-equipped laser lab~\cite{RSI_paper}, and have been completed
to the point of manufacturability~\cite{Timing_paper}. The third
and fourth development efforts have been brought to the point of
commercial manufacturability by Incom~\cite{Incom}; work continues
on improved performance and manufacturability, in particular with
respect to price and volume~\footnote{The photocathode effort has
also led to ongoing remarkable `theory-based' commercial and
accelerator development programs, beyond the scope of this talk.}
.

\begin{figure}
\includegraphics[width=.6\textwidth]{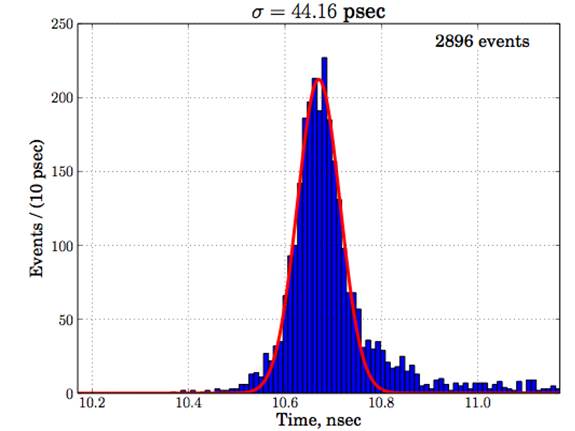}
\caption{Timing resolution for single
  photo-electrons.}
\label{fig:Timing_resolution}
\end{figure}

%Figure 8
Figure~\ref{fig:Timing_resolution} shows the measured LAPPD timing
resolution for single
  photo-electrons, digitized with the UC PSEC4 waveform sampling
  system~\cite{PSEC4_paper,Mircea_Strasbourg}.

\begin{figure}
\includegraphics[width=.6\textwidth]{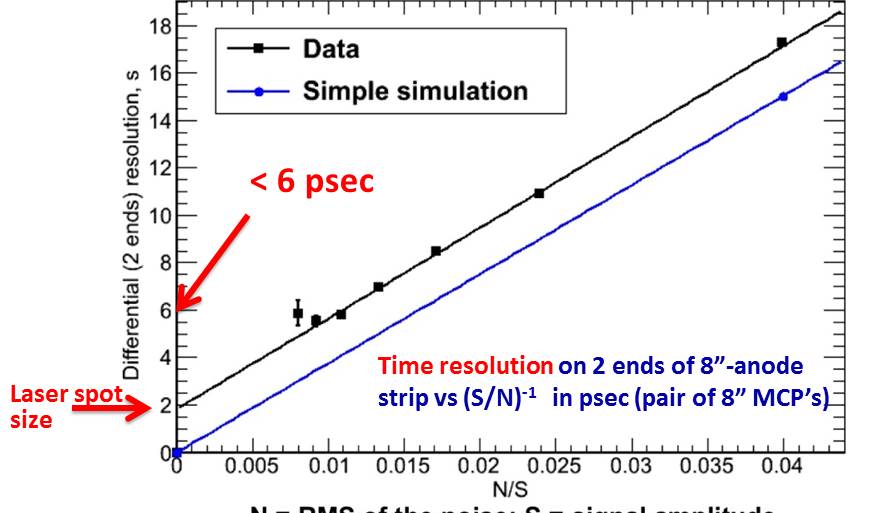}
\caption{A comparison of the measured time resolution versus the
  inverse of the Signal/Noise ratio with a simple simulation.}
\label{fig:Timing_extrapolation}
\end{figure}

Figure~\ref{fig:Timing_extrapolation} shows the measured time
resolution with the Demountable
LAPPD~\cite{Timing_paper,RSI_paper}. The simple Ritt
`rule-of-thumb' is shown for comparison. The goal of future trials
of optimized detectors in the Fermilab test beam is to test the
extrapolation down to the several psec level and below, where
there are presumably new processes that limit the resolution.

\begin{figure}
\includegraphics[width=.6\textwidth]{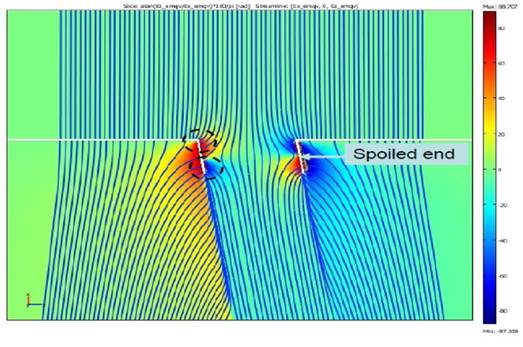}
\caption{A simulation of the electric field in a single MCP pore,
  showing the geometry of the first strike and the jitter due to path
  length differences.}
\label{fig:pore_simulation}
\end{figure}

The data are consistent with the time resolution  being dominated
by the jitter in the `first strike' of the initial photo-electron
on the secondary-emitting layer of the pore. Late in the shower
the number of electrons is very large and statistical fluctuations
are averaged out; the first several strikes, however, can vary
widely in time, dominating the transit-time spread (TTS).
 A simulation of the electric
field in an MCP pore is shown in Figure~\ref{fig:pore_simulation},
illuminating the geometry of the first strike and the jitter due
to path=length differences. The jitter due to geometry can be
diminished by smaller pore size, large bias angle, and higher
photocathode-MCP1 voltage. However to get to psec timing one will
need a large number of photo-electrons and high-enough gain so
that the first of many will initiate the leading edge of the
pulse. More simulation is needed.

\begin{figure}
\includegraphics[width=.6\textwidth]{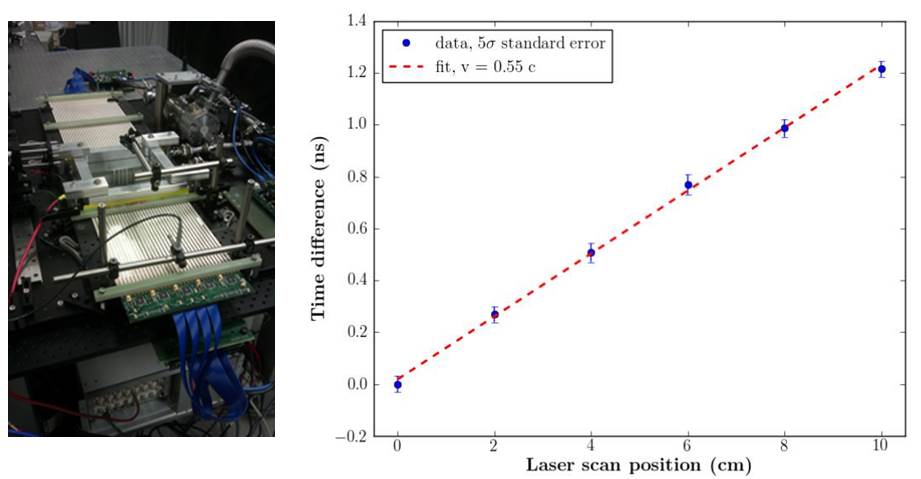}
\caption{Left: The 90-cm 4-tile RF-strip anode test station at the ANL
  APS laser lab. The PSEC4 readout is at the end of strips; the tile
  is excited by the laser on the moveable stage. Right: The measured
  time difference between the two ends of the anode versus laser
  position.}
\label{fig:4tile_anode_position_resolution}
\end{figure}

The tests of the amplification section, anode, and electronics
readout designs were done with a `Demountable' detector package
that used an pumped system with an O-ring seal, a metal
photocathode, allowing bench tests of the full performance chain
while not requiring the hermetic package and bialkali
photocathode~\cite{RSI_paper}.
Figure~\ref{fig:4tile_anode_position_resolution} shows the setup
in the Argonne APS laser lab.

%\newpage
\section{Some Proposed Applications of 1 Psec Timing for Charged
  Particles, High-Energy Gammas} 
\begin{figure}
\includegraphics[width=.6\textwidth]{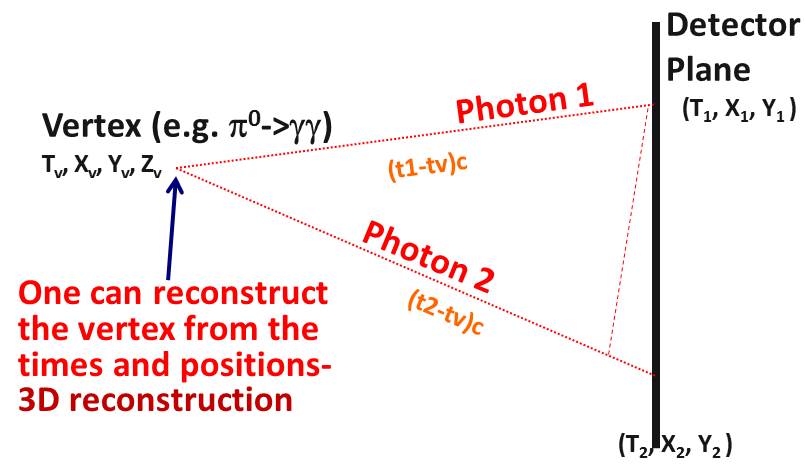}
\caption{The measurement of photonic rare decay modes of K-mesons such
the KOTO measurement of the neutral $K_L$ meson decay to pizero and neutrinos.}
\label{fig:Koto_pizero_reconstruction}
\end{figure}

The measurement of photonic rare
decay modes of K-mesons such the KOTO measurement of the neutral $K_L$
meson decay to pizero and neutrinos is dominated by combinatoric
background in which one photon comes from one $\pi^0$ or $\eta$ and a
second from another.  Figure~\ref{fig:Koto_pizero_reconstruction}
shows the over-constrained reconstruction enabled by the precise
measurement of time and position of each photon.

\begin{figure}
\includegraphics[width=.6\textwidth]{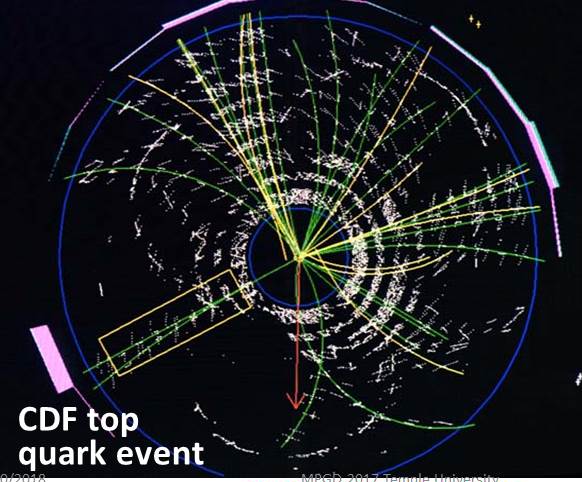}
\caption{A beams-eye view of the decay of a pair of top quarks in the
  CDF detector.}
\label{fig:CDF_top_event}
\end{figure}

At the LHC and proposed new international high-energy colliders, a
1-psec resolution would allow identification of quark flavor for each
track up to 20 GeV for s-quarks, the identification of baryons, and
the association of photons and neutrons with vertices. We have
proposed an internal `differential' time reference - the 'clock
starts' when the photons and electrons arrive at the perimeter, and
the times of arrival of heavier particles or particles from displaced
vertices are measured relative to that
zero. Figure~\ref{fig:CDF_top_event} shows a beams-eye view of the
decay of a pair of top quarks in the CDF detector at the Fermilab
Tevatron, with two W bosons decaying into (presumably) $u\bar{d}$ or
$c\bar{s}$ or their charge conjugate pairs, and a $b$ and $\bar{b}$
quark with their distinctive flavor decay chains and multiple
displaced vertices. There is a wealth of information to be had here,
and possibly also in flavor-rich rare Beyond-the-Standard-Model
signatures with low backgrounds due to highly
suppressed Standard Model diagrams.

%\newpage
\section{Beam Test of the First `Optical Time Projection Chamber'}

Time resolutions for single photons in the 10's of psec allow
using the photon drift time to reconstruct event topologies in 3
dimensions, in analogy to the similar use for electrons in
Nygren's Time Projection Chamber.  H. Nicholson named this the
`Optical Time Projection Chamber', or OTPC~\cite{Nicholson}. As
Cherenkov light is directional, and in the case of a charged
particle traversing a liquid such as water is profuse, the
technique is attractive for large liquid-based neutrino
detectors~\cite{Ypsilantis,LSND_Cherenkov,Andrey_doublebeta,Andrey_Boron8_paper,
LBL_neutrino,Minfang_Bejing_separation,Andrey_sim_talk}.

\begin{figure}
\includegraphics[width=.6\textwidth]{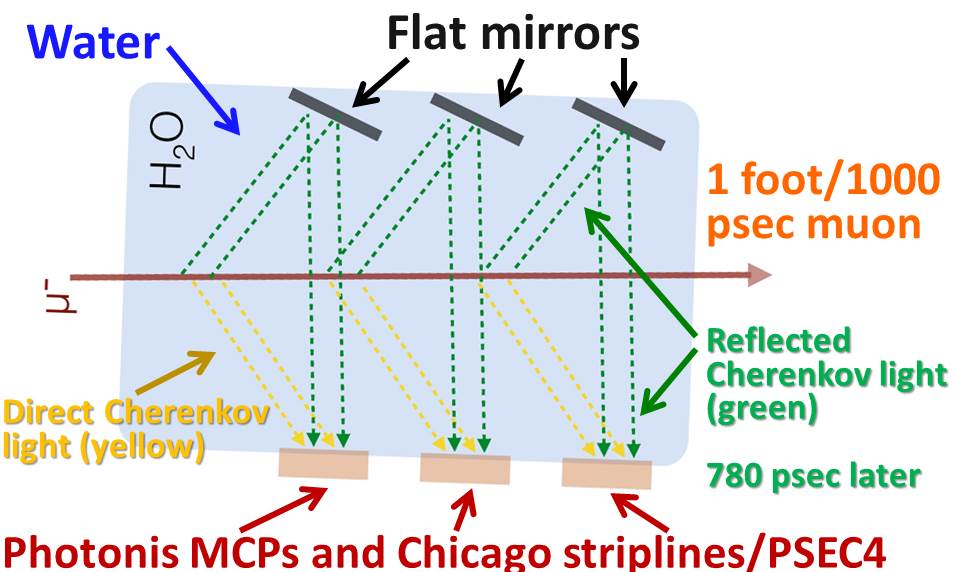}
\caption{An elevation view of Eric Oberla's design of the Optical TPC
  prototype. Note the different lengths of the optical
paths of the direct and reflected Cherenkov photons.}
\label{fig:OTPC_side_view}
\end{figure}

A small OTPC prototype has been built and the technique
demonstrated in the Fermilab Test Beam~\cite{MCenter} by E.
Oberla~\cite{OTPC_paper,Eric_thesis}. An elevation view of the
Optical TPC prototype is shown in Figure~\ref{fig:OTPC_side_view}

\begin{figure}
\includegraphics[width=.6\textwidth]{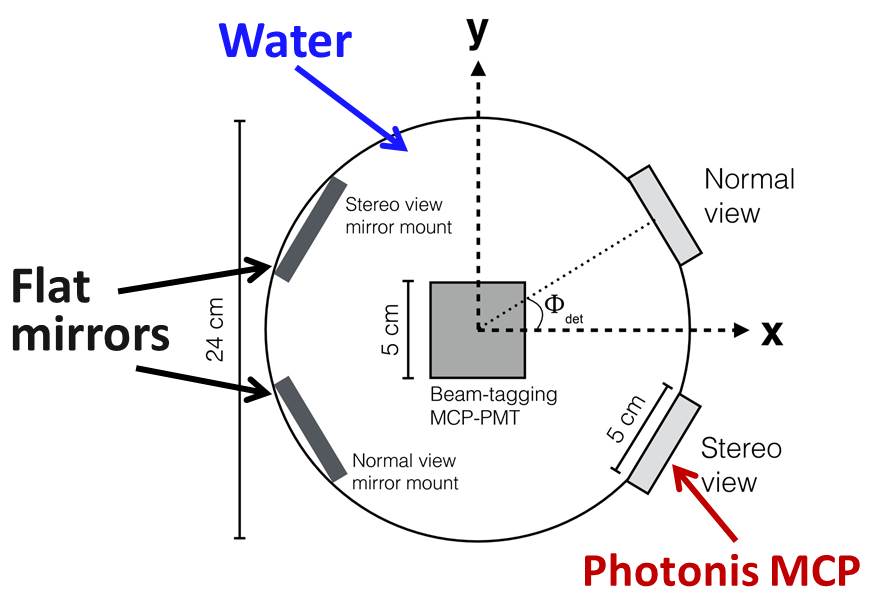}
\caption{A `beams-eye' view of the Optical TPC prototype.}
\label{fig:OTPC_beams_eye_view}
\end{figure}

The OTPC prototype recorded tracks in stereo for 3D reconstruction
with a geometry shown in Figure~\ref{fig:OTPC_beams_eye_view}

\begin{figure}
\includegraphics[width=.6\textwidth]{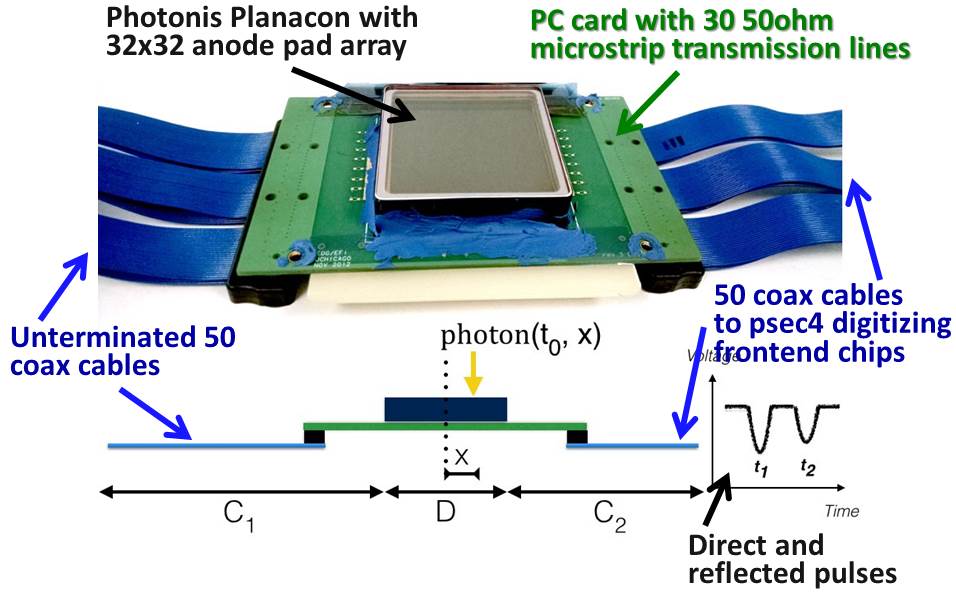}
\caption{The 50-Ohm RF strip PSEC-4 readout for the Photonis
Planacon used in the OTPC.} \label{fig:Planacon_and_strips}
\end{figure}

Oberla's implementation of 50-Ohm RF strip PSEC-4 readout for the
Photonis Planacon~\cite{Planacon} used in the OTPC is shown in
Figure~\ref{fig:Planacon_and_strips}. The anode pads were
connected~ to thirty RF-strips over the 5-cm length of the
Planacon. The readout was single-ended, recording the pulse at
both ends of the strip from just one end, by reflecting the pulse
off of an open 50-ohm line at the far end (shown as the double
pulse in the figure). This minimizes channel-to-channel
calibration systematics.

\begin{figure}
\includegraphics[width=.6\textwidth]{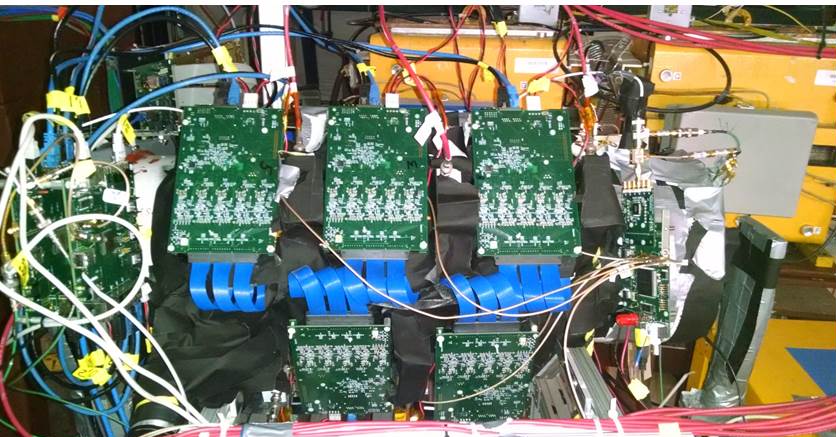}
\caption{The OTPC installed at the Fermilab Test Beam Facility.}
\label{fig:OTPC_in_beam_sideview}
\end{figure}

Figure~\ref{fig:OTPC_in_beam_sideview} shows the OTPC installed in
the beam at the Fermilab Test Beam Facility.

\begin{figure}
\includegraphics[width=.6\textwidth]{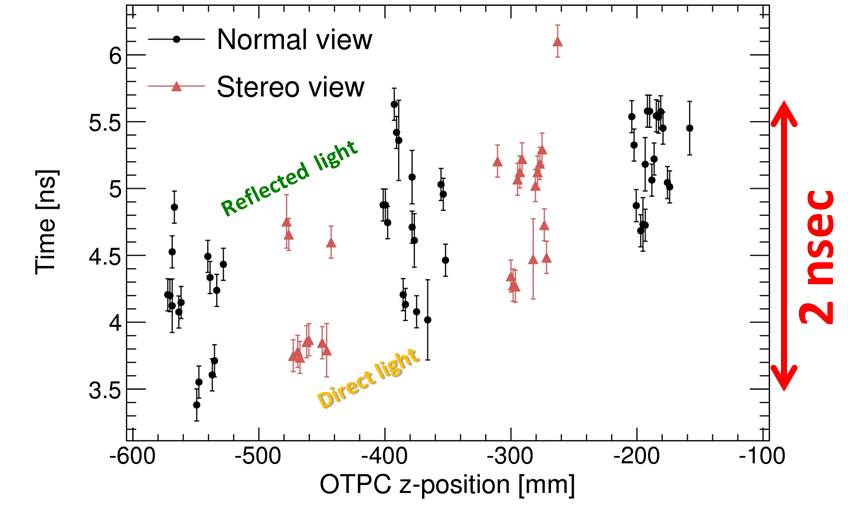}
\caption{Reconstruction of a single beam particle event in the
OTPC; the measured time of arrival of light versus distance along
the beam. The earlier light, which forms the lower track, is
direct; the reflected light, which  constitutes the upper track,
arrives 800 psec later.} \label{fig:Erico_OPTC_one_event_t_vs_z}
\end{figure}

Figure~\ref{fig:Erico_OPTC_one_event_t_vs_z} shows the measured
time versus position along the tube for a single particle (muon)
traversing the OTPC. The  angular resolution is measured to be 60
mrad over a lever arm of 40 cm~\cite{OTPC_paper,Eric_thesis}.

%\newpage
\section{New Opportunities from Exploiting Transverse-Longitudinal Phase Space}
We can apply the principle of coupled time and space arrival
coordinates for photons to localize and reconstruct images in more
complicated geometries and applications. The use of the precise
time coordinate allows working in 3D phase space, in which
mirrors, baffles, lenses, and other optical devices can mix/rotate
space and time. There are many interesting ideas to be had here.

\subsection{Search for Neutrinoless Double Beta-Decay}
\begin{figure}
\includegraphics[width=.6\textwidth]{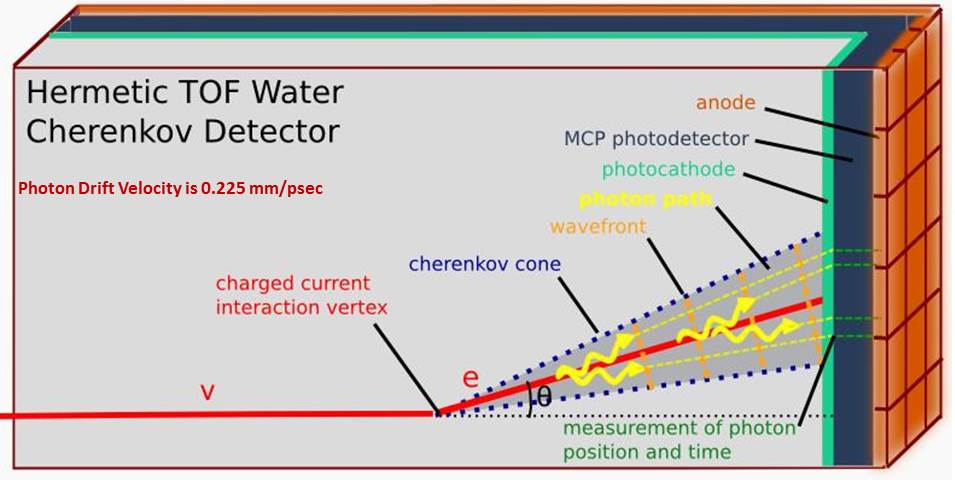}
\caption{The use of coupled time and space arrival coordinates to
  measure directionality and to fully reconstruct events in a large
  liquid neutrino detector.}
\label{fig:DUSEL_Nicholson}
\end{figure}

The process of neutrinoless double-beta decay will occur if the
neutrino is Majorana rather than Dirac. i.e. is its own
anti-particle. However the predicted rate to be tested is tiny
(typically on the order of $10^{28}$ years per nucleus, requiring
a very large number of nuclei in the examined sample.
Figure~\ref{fig:DUSEL_Nicholson} shows a sketch made in the DUSEL
era~\cite{DUSEL} showing the use of coupled time and space arrival
coordinates in reconstructing events in large liquid neutrino
detectors~\cite{Nicholson}.

\begin{figure}
\includegraphics[width=.6\textwidth]{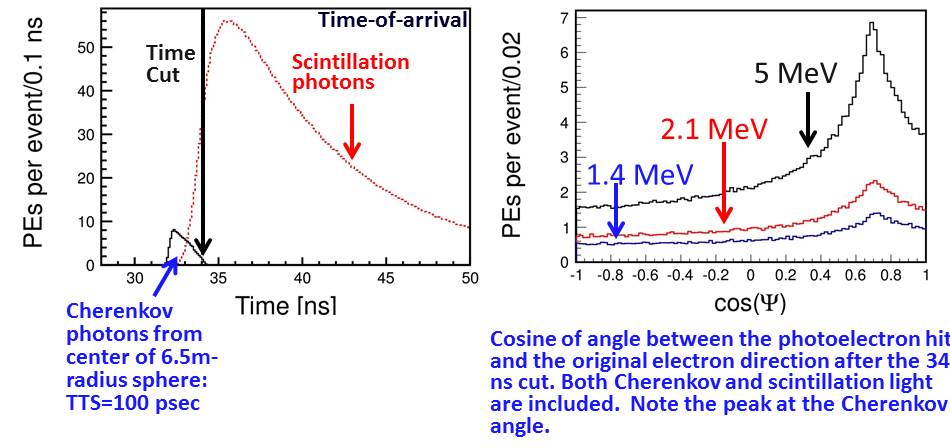}
\caption{Left: the time-of-arrival of photons from double-beta decay
  events at the center of a 6.5m radius sphere filled with water-based
  liquid scintillator. A time cut at 34 nsec enhances the Cherenkov
  light signal. Right: The cosine of the angle between the photon and
  the original electron direction, after the 34 nsec cut. Note the
  peak at the Cherenkov angle.}
\label{fig:Andrey_NLDBB}
\end{figure}

Because Cherenkov light is directional, it can be used to distinguish
the two electrons of neutrinoless double-beta decay from $^8 B$ solar
neutrinos and other backgrounds with one electron or other
signatures~\cite{Ypsilantis,Andrey_doublebeta,Andrey_Boron8_paper,Andrey_sim_talk}.
Figure~\ref{fig:Andrey_NLDBB} shows results from a simulation of
measuring directionality in neutrinoless double-beta decay in a 6.5m
radius sphere filled with water-based liquid scintillator.

The OTPC prototype mirror configuration of Figure~\ref{fig:OTPC_side_view}
multiplied up the effective photocathode coverage
by a factor of 2. One can consider going to much larger mirror/cathode
ratios (MCR) in large detectors, for which the two parameters of
coverage and MCR are major factors in cost. A significant problem is
dispersion for long path lengths, and may be mitigated by a geometry
that provides some short-path-length data for any track in addition to
more extensive coverage.

\subsection{Time-of-flight Positron Emission Tomography}

\begin{figure}
\includegraphics[width=.6\textwidth]{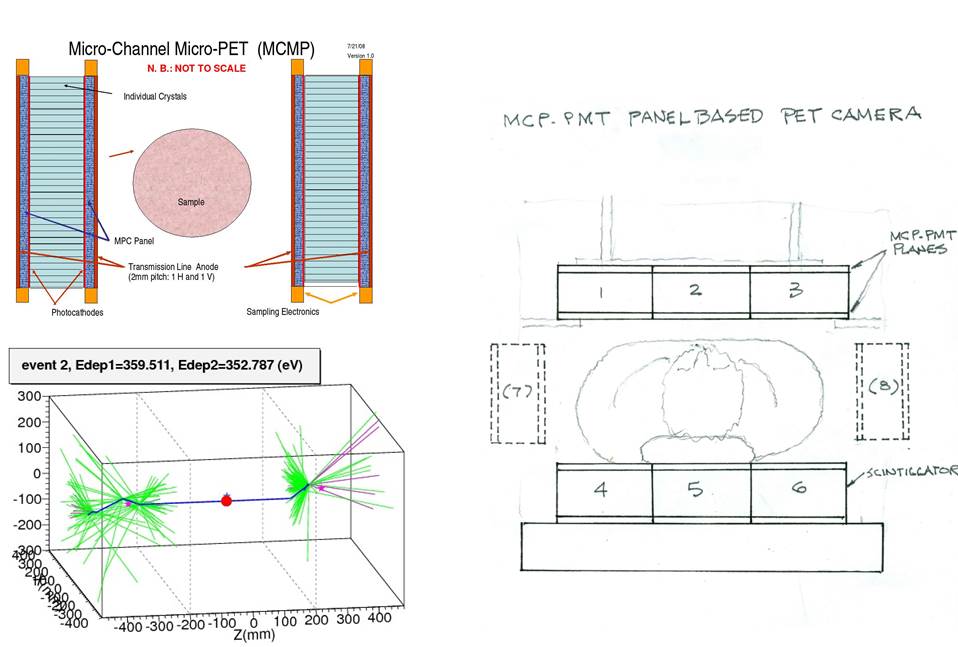}
\caption{A proposed implementation of LAPPDTM detectors  in TOF-PET
(positron-emission tomography).}
\label{fig:Pet_camera_composite}
\end{figure}

The availability of precise timing can be exploited in medical
imaging, with the goal of a whole-body low-dose TOF-PET camera for
post-diagnosis broad
localization. Figure~\ref{fig:Pet_camera_composite} show an
implementation of a TOF-PET camera that uses LAPPD modules to locate
the interactions of the two gamma rays in a liquid scintillator array.

\section{Current \LAPPDTM Status}
Incom,Inc~\cite{Incom} has established a pilot production facility for
$\LAPPDTM$ modules, and recently achieved a major milestone in
producing fully functional Generation-I (glass envelope)
$\LAPPDTM$  modules for commercial sale. These prototype
modules, while produced in low volume, are characterized by
very high gain, low noise, and useful photocathode quantum
efficiency~\cite{mjm_elba}. Several of these have recently been
sold to `early adopters', such as the ANNIE neutrino experiment at
Fermilab~\cite{ANNIE}, for characterization in specific
applications.

At UofC we are working with Incom on `Gen-II', a second generation
implementation~\cite{Andreye_elba,mjm_elba}. The mechanical footprint
is the same as for Gen-I, and the electrical properties will be the
same or better as Gen-II uses the same ALD-coated microchannel
plates. Innovations include changing the package from glass to alumina
for robustness, stability, and higher band-width; a
capacitively-coupled anode readout that moves the signal pickups
outside the vacuum volume allowing for application-specific tailoring
of a single $\LAPPDTM$ module design~\cite{InsideOut_paper}; and
possibly, PMT-like photocathode `air-transfer' batch production for
reduced cost and higher volumes!\cite{CPAD_talk}.

\section{Acknowledgements}

I thank my collaborators and technical staff at Chicago, and the
remarkable group of LAPPD collaborators at ANL, Arradiance,
Berkeley-SSL, Fermilab, Incom, and Hawaii. The participants in the
ANL- Chicago-France series of workshops made essential
contributions to the understanding of the limitations and
potential of psec timing. The work at the University of Chicago
was supported from the Department of Energy under awards DE-SC0008172 and
DE-SC0015267, the National Science Foundation under PHY-1066014, the Driskill
Foundation, and the University Physical Sciences Division. Special
thanks are due to H. Marskiske and H. Nicholson, DOE Office of
Science.

%\begin{figure}
%\includegraphics[width=.6\textwidth]{Figures/}
%\caption{}
%\label{fig:}
%\end{figure}

%===========================================================================
%===========================================================================


\begin{thebibliography}{99}
\bibitem{LAPPDTM} The University of Chicago has trademarked the LAPPD
  8''-square tile, with licensing to Incom, to distinguish it from
  smaller MCP-PMT's and fundamentally different designs.
\end{thebibliography}
\end{document}